\documentclass[prb,twocolumn,superscriptaddress]{revtex4}
\usepackage{amsmath,amssymb}
\usepackage{graphicx}
\newcommand{\be}{\begin{equation}}
\newcommand{\ee}{\end{equation}}

\begin{document}

\title{Anomalous stabilization in a spin-transfer system at high spin polarization}

\author{Inti Sodemann}

\affiliation{Department of Physics and Astronomy, University of
South Carolina, Columbia, SC 29208, USA}

\author{Ya. B. Bazaliy}

\affiliation{Department of Physics and Astronomy, University of
South Carolina, Columbia, SC 29208, USA}

\affiliation{Institute of Magnetism, National Academy of Science,
Kyiv 03142, Ukraine}

\date{\today}

\begin{abstract}
Switching diagrams of nanoscale ferromagnets driven by a
spin-transfer torque are studied in the macrospin approximation. We
consider a disk-shaped free layer with in-plane easy axis and
external magnetic field directed in-plane at 90$^{\circ}$ to that
axis. It is shown that this configuration is sensitive to the
angular dependence of the spin-transfer efficiency factor and can be
used to experimentally distinguish between different forms of
$g(\theta)$, in particular between the original Slonczewski form and
the constant $g$ approximation. The difference in switching diagrams
is especially pronounced at large spin polarizations, with the
Slonczewski case exhibiting an anomalous region.
\end{abstract}

\maketitle

\section{Introduction}
Spin polarized electric currents have been successfully used to
switch the magnetization direction of nanoscale ferromagnetic layers
via the spin transfer
effect~\cite{slonczewski,berger,tsoi,myers,sun,katine}. One of the
questions of current-induced dynamics is the dependence of
spin-transfer efficiency, or Slonczewski factor $g$, on the angle
between the polarization of incoming spin current and the
magnetization direction~\cite{slonczewski2,kovalev,xiao,zangwill}.
Such a dependence can be essential, and for example leads to the
asymmetry between the positive and negative switching currents.
However, there is still a lack of experimental tests for the precise
functional form of efficiency factor. It is expected that angular
dependence of $g$ will become more important at high spin
polarizations where the constant efficiency approximation can fail,
while constant $g$ can still be in good agreement with experimental
results at low spin polarization~\cite{wang,morise,yar,mancoff}.

Here we perform stability analysis for the equilibrium
configurations of a bilayer spin-transfer device using the
Slonczewski form for the efficiency factor and compare it with a
similar analysis that uses the constant efficiency approximation. We
observe that the switching diagram for the Slonczewski case displays
a stability region and precessional states that are absent in the
constant efficiency case. These anomalous regions become larger as
the spin polarization increases. Our results may motivate further
experimental efforts to directly measure the functional form of the
efficiency factor at high spin polarizations.

\section{Macrospin description of the device}

\begin{figure}[bp]
\begin{center}
\includegraphics[scale=0.43]{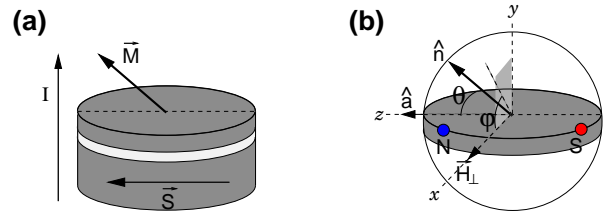}
\end{center}
\caption{\label{fig1}a) Typical nanopillar device with free layer on
the top and polarizer at the bottom. b) In-plane magnetic field
configuration. Two stable directions of the free layer magnetization
are labeled as N and S.}
\end{figure}

A typical device used to study spin-transfer effect is a nanopillar,
with two layers of ferromagnetic material separated by a normal
paramagnetic metal (see fig ~\ref{fig1}.a). The magnetization of
one layer (polarizer) is fixed and oriented along a unit vector
$\mathbf{s}$, while the magnetization of the other (free layer),
$\mathbf{M}=M \mathbf{n}$, rotates and is described in the macrospin
approximation by the Landau-Lifshitz-Gilbert (LLG) equation
including the Slonczewski spin torque term~\cite{slonczewski}
\begin{equation}\label{LLG}
\mathbf{\dot{n}}=\frac{\gamma}{M}\Bigg[-\frac{\delta E}{\delta \mathbf{n}}\times
\mathbf{n}\Bigg]+\frac{\gamma \hbar I}{2 e V M}
g(\theta,P) [\mathbf{n}\times(\mathbf{s}\times
\mathbf{n})]+\alpha[\mathbf{n}\times\mathbf{\dot{n}}],
\end{equation}
where $\gamma$ is the gyromagnetic ratio, $E(\mathbf{n})$ is the
magnetic energy of the free layer, and $\alpha$ is the Gilbert
damping constant. The strength of the spin-torque is characterized
by the efficiency factor, $g(\theta,P)$, which depends on the angle
$\theta$ between the magnetizations of the polarizer and the free
layer, and the degree of current spin polarization $P\in[0,1]$. In
general the functional form of $g(\theta,P)$ is material and
geometry-dependent~\cite{slonczewski2,kovalev,xiao,zangwill}. Here
we will compare Slonczewski's~\cite{slonczewski} form
 \be
g(\theta,P)=\frac{1}{f_P(\xi_P+\cos \theta)},
 \ee
with $\xi_P=3-4/f_P$, $f_P=(1+P)^3/4P^{3/2}$ and the
$g(\theta,P)$=const approximation.

The magnetic energy of the free layer includes contributions from
the intrinsic anisotropy (easy axis anisotropy with strength $H_a$
and direction $\mathbf{\hat{a}}$), shape anisotropy (easy plane
anisotropy with normal vector $\mathbf{\hat{p}}$) and the
interaction energy with an external magnetic field $\mathbf{H}$.
Equation \eqref{LLG} can be written as
 \be\label{sLLG}
 \mathbf{\dot{n}}=\boldsymbol{\tau}(\mathbf{n})+\alpha
 \mathbf{n}\times \boldsymbol{\tau}(\mathbf{n}) \ ,
 \ee
where we have rescaled the time as $T= t/(1+\alpha^2)$, and $\boldsymbol{\tau}$
is defined as
\be
\begin{aligned}\label{vectau}
\boldsymbol{\tau}(\mathbf{n})&=-\nabla \varepsilon(\mathbf{n})\times \mathbf{n}+\omega_I g(\theta,P) \
\mathbf{n}\times(\mathbf{s}\times\mathbf{n}), \\
\varepsilon(\mathbf{n})&=\frac{\omega_p}{2}(\mathbf{\hat{p}\cdot\mathbf{n}})^2-\frac{\omega_a}{2}(\mathbf{\hat{a}\cdot\mathbf{n}})^2-\omega_H(\mathbf{\hat{h}}\cdot\mathbf{n}).
\end{aligned}
\ee
The newly defined constants are related to the already introduced
parameters according to
 \be
\begin{aligned}
\omega_a&=\gamma H_a,&   \omega_p& =4\pi \gamma M, \\
\omega_H& =\gamma H, &   \omega_I&=\frac{\gamma \hbar}{2eVM}I.
\end{aligned}
 \ee
All of them have dimensions of frequency making the comparison
between the terms of different origin straightforward. In accord
with experimental situations it is assumed that
$\omega_I\ll\omega_p$.

We study a device with an in-plane easy axis and the in-plane
magnetic field perpendicular to it (see fig.~\ref{fig1}.b). Choosing
the system of coordinates
$\mathbf{\hat{s}}=\mathbf{\hat{a}}=\mathbf{\hat{e}_z}$,
$\mathbf{\hat{h}}=\mathbf{\hat{e}_x}$,
$\mathbf{\hat{p}}=\mathbf{\hat{e}_y}$, we obtain, from equation
\eqref{vectau}, the components of $\boldsymbol{\tau}$ in spherical
coordinates
 \be\label{tau}
\begin{split}
\tau_{\phi}&=\frac{1}{2}\sin2\theta(\omega_p\sin^2\phi+\omega_a)-\omega_H\cos\theta\cos\phi,\\
\tau_{\theta}=&-\frac{\omega_p}{2}\sin\theta\sin2\phi-\omega_H\sin\phi-\omega_Ig(\cos\theta)\sin\theta.
\end{split}
 \ee
The equilibrium directions of the magnetization $\bf n$ correspond
to the solutions of the equation $\boldsymbol{\tau}(\mathbf{n})=0$. Here we
consider the two in-plane equilibrium points. At $\omega_I = 0$,
$\omega_H = 0$ these are the north (N) and the south (S) poles. For
$\omega_I =0$, $\omega_H \neq 0$ they shift and approach the
direction of magnetic field, finally merging at $\omega_H =
\omega_a$. The shifted equilibrium points are still labeled by N and
S (Fig.~\ref{fig1}b).

Stability of an equilibrium can be checked by expanding
$\mathbf{\tau}$ in angular deviations $\delta\theta$, $\delta\phi$,
and writing equation \eqref{sLLG} in an approximate form

\be
\dbinom{\dot{\phi}}{\dot{\theta}}=\mathbf{D}\dbinom{\delta \phi}{\delta \theta}.
\ee

The equilibrium is stable when the real parts of both eigenvalues of
$\mathbf{D}$ are negative, or equivalently when matrix $\bf D$
satisfies ${\rm Tr}\mathbf{D}<0$ and ${\det}\mathbf{D}>0$ at the
equilibrium.

\begin{figure}
\includegraphics[scale=0.75]{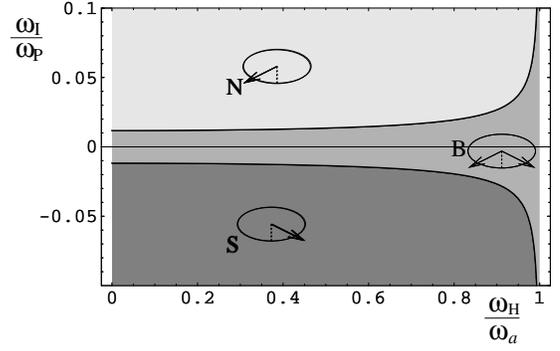}
\caption{\label{fig2} Switching diagram for the $g = $const
approximation. The value of $g$ is chosen as the average value of
Slonczewski's $g(\theta)$ used in Fig.~\ref{fig3}. Other parameters
are set to $\omega_a/\omega_p=0.01$, $P=0.7$, and $\alpha=0.01$.
Stability regions for the N and S equilibria (see text) overlap
forming the bistable region marked as B.}
\end{figure}

\section{Stability regions}

The modified positions of the N and S equilibria for $\omega_H \neq
0$, $\omega_I \neq 0$ are given by
 \be
\begin{split}
\label{theta} \sin\theta_{N,S}
 & = \frac{\omega_H}{\omega_a}+\mathcal{O}\left(\frac{\omega_I}{\omega_P}\right)^2,
 \\ \sin\phi_{N,S}=
 - & g_{N,S}  \frac{\omega_I}{\omega_p}\left(1+\frac{\omega_a}{\omega_p}\right)
 +\mathcal{O}\left(\frac{\omega_I}{\omega_P}\right)^2,
\end{split} 
 \ee
for $0< \omega_H < \omega_a$ (with $g_{N,S}=g(\theta_{N,S})$). The
angle $\theta$ and the magnetic field strength $\omega_H$ have a
one-to-one correspondence and can be used interchangeably. The trace
of $\mathbf{D}$-matrix at these points can be found as
 \be
\begin{split}
\rm Tr \mathbf{D}=&-\omega_I\left[g(\theta)\cos\theta+\frac{d}{d\theta}(g(\theta)\sin\theta)\right]\\&-\alpha[\omega_p\cos2\phi+(\omega_p\sin^2\phi+\omega_a)(1+\cos^2\theta)]
\end{split}
 \ee
Approximations \eqref{theta} give the stability condition in the
form
 \be
 \begin{split}
 \alpha\left[1+\frac{\omega_a}{\omega_p}(1+\cos^2\theta)\right]> -&\frac{\omega_I}{\omega_p}\left[2g(\theta)\cos\theta+g'(\theta)\sin\theta\right]\\&+\mathcal{O}\left(\frac{\omega_I}{\omega_P}\right)^2.
 \end{split}
 \ee
The determinant
 \be
\frac{1}{1+\alpha^2}\det \mathbf{D}=\omega_a(\omega_p+\omega_a)\cos^2\theta+\mathcal{O}(\omega_I^2),
 \ee
in the small current regime remains positive, so it does not play
any role in the stability analysis in this case. In contrast, the
trace ${\rm Tr}\mathbf{D}$ is more sensitive and can change sign as
the current is varied. Moreover, the explicit appearance of
$g'(\theta)$ in the formula leads to important differences in the
switching diagrams for different forms of $g(\theta)$. In the case
of constant $g$-factor the stability condition for N- and
S-equilibira can be written as
 \be
\omega_I\gtrless \mp \alpha
\frac{\omega_p+\omega_a(2-(\omega_H/\omega_a)^2)}{2g\sqrt{1-(\omega_H/\omega_a)^2}}
 \ee
where $>$, $-$ ($<$, $+$) corresponds to the N (S) stability region
(See Fig.~\ref{fig2}). The switching current exhibits the $1/\cos
\theta$ divergence reported in the experiments for this
regime~\cite{mancoff}.

\begin{figure}
\includegraphics[scale=0.75]{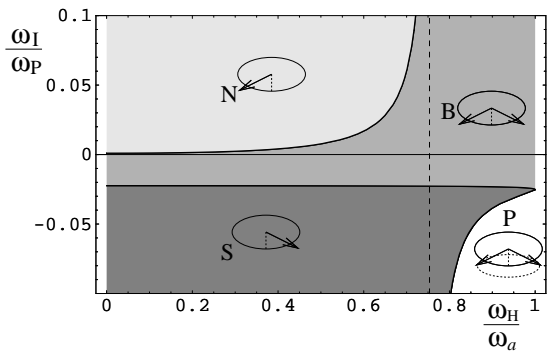}
\caption{\label{fig3} Switching diagrams for the Slonczewski's form
of $g$-factor. Other parameters are the same as in Fig.~\ref{fig2}.
Regions of stability for north (N) and south (S) poles overlapping
in the bistable region (B), and the region of precessional states
(P) are shown. The onset of the anomalous stability behavior occurs
at a field $\omega^*_H/\omega_P = 0.76$.}
\end{figure}

For the Slonczewski $g$-factor, the condition of stability for the N
point is

\begin{equation}
\omega_I> -\alpha
\frac{\omega_p+\omega_a(2-(\omega_H/\omega_a)^2)}{2g_{N}\sqrt{1-(\omega_H/\omega_a)^2}+
g^2_{N} f_P (\omega_H/\omega_a)^2},
\end{equation}
 
whereas for the S point the condition becomes

\begin{equation}\label{South}
\omega_I \lessgtr \alpha
\frac{\omega_p+\omega_a(2-(\omega_H/\omega_a)^2)}{2g_{S}\sqrt{1-(\omega_H/\omega_a)^2}-
g^2_{S} f_P (\omega_H/\omega_a)^2},
\end{equation}

where $<$ ($>$) is the condition for $\omega < \omega^*_H$ ($\omega
> \omega^*_H$), $\omega^*_H$
designates the field for which the denominator of equation
\eqref{South} becomes zero, and determines the onset of an stability
behavior completely absent in the $g$-constant case
(Fig.~\ref{fig3}). This field, or equivalently the angle
characterizing the S point, depends only on polarization $P$ and can
be found from
 
\be
 \cos\theta_{c} =
 -\sqrt{1-(\omega_H^*/\omega_a)^2}=\sqrt{\xi^2_P-1}-\xi_P.
\ee

\begin{figure}
\begin{center}
\includegraphics[scale=0.7]{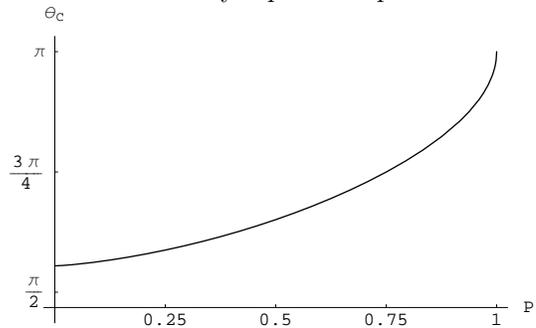}
\end{center}
\caption{\label{fig4} Critical angle $\theta_c$ for the onset of the
anomalous stabilization as function of the polarization}
\end{figure}

In the ``anomalous'' regime $\omega_H > \omega_H^*$ a current of
positive polarity stabilizes both N and S points, while the
application of a sufficiently large negative current destabilizes
both points (see Fig.~\ref{fig3}), moreover, there is a region in
which none of the equilibria are stable, suggesting the existence of
precessional motion. The value of $\omega^*_H$ becomes smaller as
the polarization increases. In the limit $P \rightarrow 1$ it
becomes zero, so that the anomalous region fills all the
switching diagram. In other words, as the polarization becomes
larger the differences between the $g$-constant approximation and the
Slonczewski form become quite dramatic. The position $\theta_c$ of
the S point at $\omega_H^*$ is shown in Fig.~\ref{fig4}.

Substantial difference between the switching diagrams at large spin
polarizations found in this study underscores the necessity of
developing new experiments capable of determining the $g(\theta)$
dependence. It also suggests that in the regime of large spin
polarization the behavior of spin-transfer devices may experience
qualitative changes.

\begin{acknowledgments}
The authors are grateful to S. Garzon for many stimulating
discussions.
\end{acknowledgments}

\end{document}